\documentclass[11pt,a4paper]{article}

\usepackage[utf8]{inputenc}
\usepackage[T1]{fontenc}
\usepackage{amsmath}
\usepackage{amssymb}
\usepackage{graphicx}
\usepackage{hyperref}
\usepackage{geometry}
\usepackage{setspace}
\usepackage{authblk}
\usepackage{booktabs}
\usepackage{caption}
\usepackage{subcaption}
\usepackage{xcolor}
\usepackage[numbers,sort&compress]{natbib}
\usepackage{titlesec}
\usepackage{enumitem}
\usepackage{float}
\usepackage{array}
\usepackage{multirow}
\usepackage{url}
\usepackage{lineno}
\usepackage{tabularx}
\usepackage{longtable}
\usepackage{makecell}
\usepackage[labelfont=bf,labelsep=period]{caption}
\hypersetup{
    colorlinks=true,
    linkcolor=blue,
    citecolor=blue,
    urlcolor=blue
}
\renewenvironment{abstract}{%
  \small
  \begin{center}
    {\bfseries \abstractname}
  \end{center}
  \quote
}{%
  \endquote
}

\titleformat{\section}{\large\bfseries}{\thesection.}{0.5em}{}
\titleformat{\subsection}{\normalsize\bfseries}{\thesubsection.}{0.5em}{}
\titleformat{\subsubsection}{\normalsize\itshape}{\thesubsubsection.}{0.5em}{}

\title{%
\rule{\textwidth}{2pt}\\

{\LARGE\textbf{DrugGen 2: A disease-aware language model for enhancing drug discovery}}\\[0.5ex]
\rule{\textwidth}{2pt}
}

\author{%
    \large
    Ali Motahharynia$^{1,2, \dagger, \star}$,\;
    Mohammadreza Ghaffarzadeh-Esfahani$^{1,\dagger}$,\;
    Mahsa Sheikholeslami$^{1, 3}$,\;
    Navid Mazrouei$^{1}$,\;
    Matin Irajpour$^{1, 4}$,\
    Yousof Gheisari$^{1, 5}$,\;
    Hajar Sirous$^{6}$,\;

    \small\itshape
    $^{1}$Regenerative Medicine Research Center, Isfahan University of Medical Sciences, Isfahan, Iran\\
    \small\itshape
    $^{2}$Isfahan Neuroscience Research Center, Isfahan University of Medical Sciences, Isfahan, Iran\\
    \small\itshape
    $^{3}$Department of Medicinal Chemistry, School of Pharmacy, Isfahan University of Medical Sciences, Isfahan, Iran\\
    \small\itshape
    $^{4}$Isfahan Cardiovascular Research Center, Cardiovascular Research Institute, Isfahan University of Medical Sciences, Isfahan, Iran\\
    \small\itshape
    $^{5}$Department of Genetics and Molecular Biology, Isfahan University of Medical Sciences, Isfahan, Iran\\
    \small\itshape
    $^{6}$Bioinformatics Research Center, Isfahan University of Medical Sciences, Isfahan, Iran\\

\bigskip
    \small\normalfont
    $^{\dagger}$These authors contributed equally to this work\\

    $^{\star}$Correspondence:\\
    Ali Motahharynia:
    \href{mailto:alimotahharynia@gmail.com}{alimotahharynia@gmail.com},\;
    Tel/Fax: +98-3136687087,\;
    ORCID: 0000-0002-1140-3257\\
}
\date{\rule{\textwidth}{0.4pt}}

\begin{document}

\maketitle

% ─── Abstract ───────────────────────────────────────────────────────────────
\begin{abstract}

Current computational approaches for drug design typically focus on generating molecules conditioned on specific targets or general molecular properties, often neglecting the influence of disease context on target behavior and therapeutic outcomes. To address this gap, we introduce DrugGen-2, a novel generative model that designs small molecules conditioned on both disease ontology and target protein sequences. DrugGen-2 was developed by fine-tuning a pre-trained GPT-2 model on a curated dataset of approved drugs linked to their diseases and targets, using a two-step strategy of supervised fine-tuning followed by reinforcement learning via group relative policy optimization (GRPO). This process was guided by reward functions optimizing for chemical validity, novelty, diversity, and high predicted binding affinity. When evaluated on five protein targets relevant to diabetic nephropathy, DrugGen-2 significantly outperformed baseline models (DrugGPT and DrugGen). It demonstrated a superior capacity to generate unique molecules, exhibited greater structural similarity to approved drugs, and achieved improved predicted binding affinities across all targets. Molecular docking analyses further supported these findings, identifying candidate ligands with strong binding potential, including compounds with predicted affinities (--9.917, --9.485, and --9.367) exceeding those of reference drugs such as enalapril for angiotensin-converting enzyme (--8.283). By integrating disease-specific context into molecular generation, DrugGen-2 advances AI-assisted drug discovery, offering a powerful tool for de novo design and drug repurposing that accounts for the complex interplay between diseases and molecular targets.

\vspace{0.5cm}
\noindent\textbf{Keywords:} Drug design; Drug repositioning; Large language model; Reinforcement learning
\end{abstract}

% ─── Introduction ───────────────────────────────────────────────────────────
\newpage
\section{Introduction}

Drug discovery is a multifaceted and resource-intensive process that involves multiple stages, from the design of high-quality small molecules to clinical translation~\cite{ref1}. Traditionally, classical approaches such as phenotypic screening and target-based drug design have been used, wherein compounds were either screened for biological activity or synthesized to interact with specific molecular targets, but these strategies remain limited in their ability to efficiently explore the vast chemical space ~\cite{ref2, ref3}. Computational methods and machine learning (ML) have begun to address these challenges by enabling more accurate prediction of molecular properties, biological activity, and synthetic feasibility ~\cite{ref4}. More recently, advances in deep learning, especially generative models, have further changed the field, allowing the design of novel molecules with improved efficiency by learning complex biological and chemical patterns ~\cite{ref5}.
Among generative approaches, large language models (LLMs) stand out for their capacity to process sequential molecular representations to capture subtle patterns in structure--property relationships ~\cite{ref6}. Examples are models such as Mol-LLM ~\cite{ref7}, Token-Mol ~\cite{ref8}, and DrugLLM ~\cite{ref9} that integrate multi-modal information for tasks including property prediction and conformation generation. When applied specifically to drug generation, these LLMs have facilitated the creation of antiviral candidates, such as TransAntivirus ~\cite{ref10}, as well as target-aware molecules exemplified by TamGen ~\cite{ref11} and DrugGPT ~\cite{ref12}. They have also facilitated ligand optimization through reinforcement learning, as demonstrated by DrugGen ~\cite{ref13}, which incorporates feedback mechanisms to enhance validity, novelty, and binding affinity.
Despite these advancements, a notable limitation remains: most existing approaches condition molecular generation on protein targets or general molecular properties without tailoring generation to specific diseases, leaving a gap in disease-conditioned drug design. This limits their applicability in personalized therapeutic development, where the same target may behave differently across disease states, or where specific disease-associated pathways require context-specific molecular design ~\cite{ref14, ref15}. For example, peroxisome proliferator-activated receptor gamma (PPAR$\gamma$) activation exerts beneficial metabolic effects in type 2 diabetes but can drive divergent outcomes in colon cancer due to disease-specific interactions with the Wnt/$\beta$-catenin pathway and metabolic state ~\cite{ref16}.
To fill this gap, we developed DrugGen-2, an extension of our previously reported DrugGen framework, designed to generate molecules conditioned on both disease ontology (medical subject headings (MeSH)) and target sequence. We developed DrugGen-2 by fine-tuning DrugGPT, a pretrained GPT-2 model specialized for ligand generation, on a curated dataset of approved drugs and their related disease-target information. The training strategy involved two complementary steps: supervised fine-tuning (SFT) ~\cite{ref17} to adapt the model to the domain-specific data, followed by group relative policy optimization (GRPO) ~\cite{ref18}, a reinforcement learning approach designed to enhance the model's generative capabilities for the generation of molecules that are chemically valid, structurally novel with high binding affinity, and similar to known approved drugs. Docking evaluations further confirm DrugGen-2's ability to produce biologically relevant small molecules. By accepting a disease MeSH directed acyclic graph (DAG) and a target amino acid sequence as inputs, DrugGen-2 outputs high-quality simplified molecular input line entry system (SMILES) for that disease--target pair, offering a platform for both drug repurposing and de novo design, paving the way toward disease-aware drug discovery.

% ─── Results ────────────────────────────────────────────────────────────────
\section{Results}

\begin{figure}[!ht]
\centering
\includegraphics[width=0.8\textwidth]{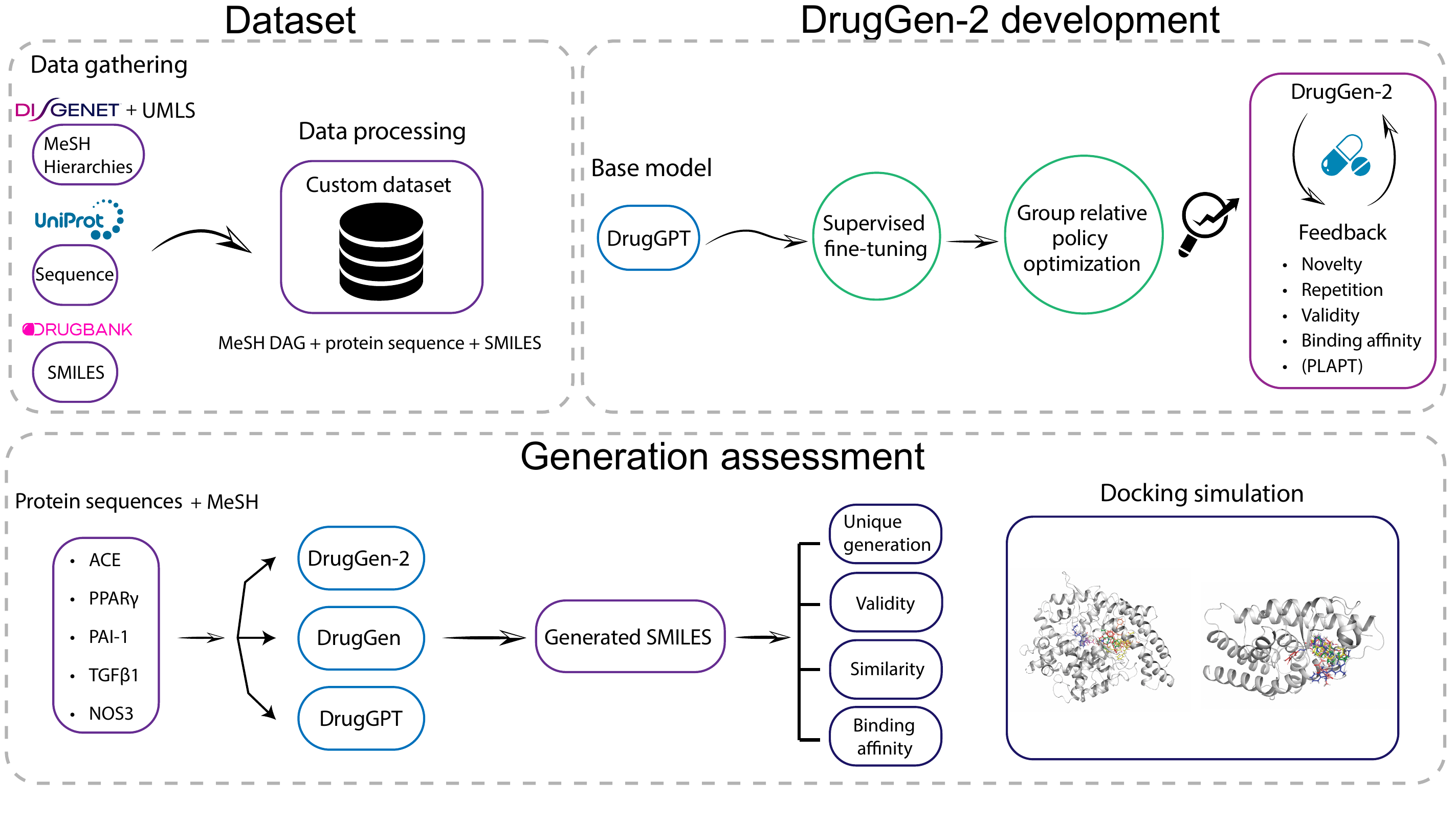} 
 \caption{\textbf{Schematic representation of DrugGen-2 development and evaluation.} The top section shows dataset preparation and the training process through supervised fine-tuning (SFT) and group relative policy optimization (GRPO) using a customized reward function. The bottom section outlines the assessment process, based on unique generation, validity, similarity to approved drugs, and binding affinity for DrugGen-2, DrugGen, and DrugGPT, along with docking simulations for DrugGen-2.}
\label{fig:1}
\end{figure}

We developed DrugGen-2 using a curated approved disease--target--drug dataset through a two-step training strategy: SFT followed by GRPO (Figure~\ref{fig:1}). In the supervised phase, the model was aligned with disease--target--drug relationships, while the reinforcement learning phase optimized generation toward producing chemically valid, novel, diverse, and high-affinity molecules. Because DrugGen-2 accepts MeSH DAG numbers as input, and diabetic nephropathy (DN) is annotated with four such identifiers, we evaluated performance separately for each entry. For benchmarking, we focused on five protein targets associated with DN: angiotensin-converting enzyme (ACE), PPAR$\gamma$, nitric oxide synthase 3 (NOS3), plasminogen activator inhibitor-1 (PAI-1), and transforming growth factor beta 1 (TGF-$\beta$1), as identified by DisGeNET ~\cite{ref19} and DrugTar algorithm ~\cite{ref20}. Outputs from DrugGen-2 were then compared with those generated by DrugGPT and DrugGen across multiple evaluation criteria, including generation capacity, structural validity, similarity to approved molecules, and predicted binding affinity.

\subsection{Reward functions efficiently guided DrugGen-2 during training}

\begin{figure}[!ht]
\centering
\includegraphics[width=0.8\textwidth]{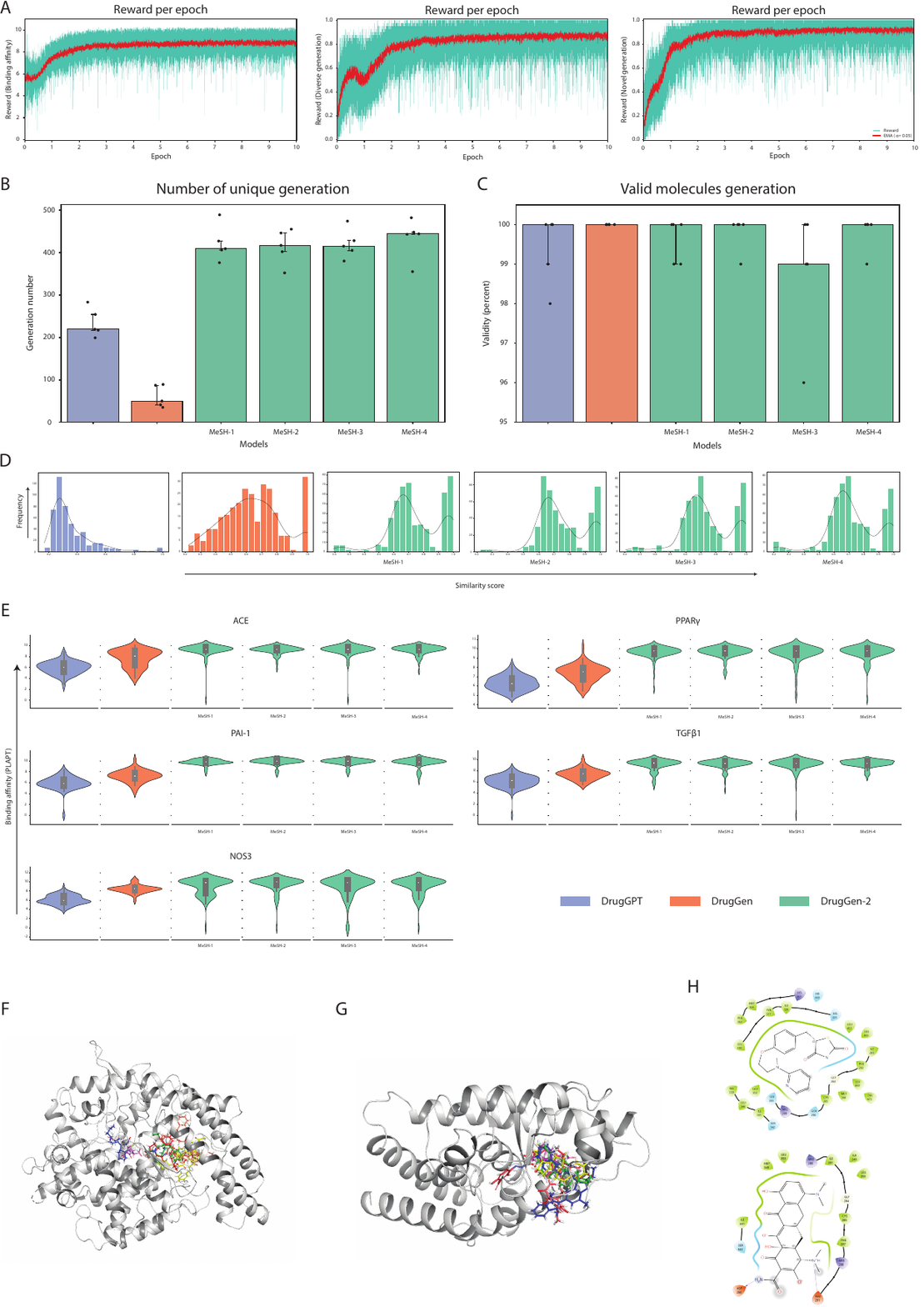} 
\caption{\textbf{Evaluation of the DrugGen-2 algorithm.} 
\textbf{(A)}Reward progression during group relative policy optimization (GRPO). Comparison of DrugGen-2 with DrugGen and DrugGPT in terms of  
\textbf{(B)} number of unique generations, 
\textbf{(C)} valid generations, 
\textbf{(D)} similarity to approved molecules, and 
\textbf{(E)} predicted binding affinity. 
\textbf{(F)} Visualization of protein--ligand complexes at the active site of 1UFZ (P12821), showing the binding of Enalapril (red), ligands P12821-293 (orange), P12821-10 (blue), P12821-269 (green), P12821-225 (yellow), and Captopril (pink).
\textbf{(G)} Binding site of 4EMA (P37231) displaying Balsalazide (red), Rosiglitazone (pink), and ligands P37231-165 (orange), P37231-117 (blue), P37231-39 (green), and P37231-247 (yellow). 
\textbf{(H)} Visualization of 3D protein-ligand interactions, Rosiglitazone (top) and P37231-165 (bottom) in PPAR$\gamma$'s active site.
}
\label{fig:2}
\end{figure}

In the training process, after six epochs of supervised fine-tuning using the SFT method, GRPO was employed as a reinforcement learning technique to further refine the model. Three reward functions were designed to guide this optimization process: (i) binding affinity, evaluated using the deep learning model, protein ligand binding affinity prediction using pre-trained transformers (PLAPT)~\cite{ref21}, in combination with a customized invalid structure assessor, (ii) molecular diversity within each training batch, and (iii) novelty of generated molecules compared to approved drugs. The training was conducted for 10 epochs, during which all reward functions gradually converged, indicating a stable optimization process. (Figure~\ref{fig:2}A and Supplementary File 1).

\subsection{DrugGen-2 achieves higher unique generation across targets compared to DrugGen and DrugGPT}

We assessed each model's capacity to generate unique molecules by tasking them to produce 500 unique candidates for the selected protein targets (Supporting information 2). Across all targets and MeSH configurations, DrugGen-2 consistently achieved the highest number of unique generations, ranging from 409 [406--427] to 444 [443--448] compared to DrugGen (50 [41--87]) and DrugGPT (219 [217--254]) ($\chi^2$ = 88.80, P < 10-9, Cram\'er's V = 0.05, Fig.~\ref{fig:2}B, Supporting information 3-Table 1-2). These results demonstrate that DrugGen-2 possesses a markedly greater capacity to generate unique molecules across targets, with consistent performance across different MeSH DAG inputs.

\subsection{DrugGen-2 produces highly valid small molecules}

We evaluated the structural and chemical validity of generated molecules using a customized validity assessor. For this analysis, we tasked each model to generate 100 unique molecules for each target (DrugGPT and DrugGen) or per MeSH-target pair (DrugGen-2) (Supporting information 2). DrugGen-2 achieved near-perfect performance on generating valid small molecules across all five targets, with median values ranging from 99 to 100% across MeSH hierarchies (Fig. 2C and Supporting information 3-Table 3). Validity was comparable to DrugGPT and DrugGen, with no significant differences observed ($\chi^2$ = 0.12, P = 1.0; Cram\'er's V = 0.003).

\subsection{DrugGen-2 demonstrates strong similarity to approved drugs}

To assess the similarity between generated molecules and approved drugs, the same set of small molecules generated in the validity assessment was used. DrugGen-2 generated molecules with higher similarity to approved drugs across all MeSH categories (0.70), compared to DrugGPT (0.30 [0.26--0.41]) and DrugGen (0.64 [0.51--0.76]) (H = 961.88, $\epsilon^2$ = 0.34, P < 10-204, Fig.~\ref{fig:2}D and Supporting information 2 and 3-Table 4-5). No significant differences were observed among the MeSH-specific DrugGen-2 variants. These findings indicate that DrugGen-2 produces molecules structurally closer to approved drugs, highlighting its potential in drug discovery.

\subsection{DrugGen-2 demonstrates improved binding affinity across five key targets}

The same set of molecules generated in the previous assessment was used to predict the binding affinity using PLAPT (21). DrugGen-2 consistently outperformed DrugGPT and DrugGen across all five DN-associated targets (Fig.~\ref{fig:2}E, Supporting information 2 and 3-Table 6). Median affinities for DrugGen-2 variants ranged from 9.26 to 9.97, markedly higher than those of DrugGPT (5.86 to 6.22) and DrugGen (7.15 to 8.49). These improvements were highly significant across all targets (ACE: $\epsilon^2$ = 0.49, P < 10-59, PAI-1: $\epsilon^2$ = 0.57, P < 10-65, PPAR$\gamma$: $\epsilon^2$ = 0.57, P < 10-67, TGFB1: $\epsilon^2$ = 0.50, P < 10-55, NOS3: $\epsilon^2$ = 0.34, P < 10-38, Supporting information-3 Tables 7-11). No significant differences were observed among MeSH-specific DrugGen-2 variants, indicating consistent performance regardless of disease ontology encoding. Overall, DrugGen-2 exhibits enhanced molecular optimization, providing superior predicted binding across multiple targets.

\subsection{Docking simulation determined the quality of the generated molecules}

To evaluate the binding interactions of molecules generated by DrugGen-2, two proteins with available crystal structures from the protein data bank (PDB), ACE and PPAR$\gamma$, were selected. A total of 125 molecules were generated for each MeSH--target pair and subjected to docking analysis (Supporting information 4). Docking was performed using the GLIDE extra precision (XP) scoring function. Several designed ligands achieved docking scores comparable to or better than their reference drugs, suggesting enhanced binding potential (Table~\ref{tab:1} and Supporting information 4). For ACE, compounds P12821-293, P12821-10, and P12821-269 showed markedly lower docking scores than Enalapril (--9.917, --9.485, and --9.367 vs. --8.283, respectively; Table 1), indicating stronger predicted affinities. Moreover, their binding poses within the ACE active site showed close spatial alignment with the reference ligand (Figure~\ref{fig:2}F). By contrast, none of the P37231 ligands surpassed Rosiglitazone or Balsalazide in docking score against PPAR$\gamma$ (Figure~\ref{fig:2}G and Table 1), although some displayed unique pharmacophoric interactions within the active site. Figure~\ref{fig:2}H highlights the binding mode of P37231-165 in PPAR$\gamma$, which introduces novel substructures absent in Rosiglitazone and may form new optimizable interactions despite its lower overall docking score.
To assess the reliability of the docking protocol, co-crystallized ligands were re-docked into their protein targets and root-mean-square deviation (RMSD) values calculated between experimental and predicted poses. For Rosiglitazone docked into PPAR$\gamma$ (PDB ID: 4EMA), the RMSD was 1.44 \AA{}, indicating faithful reproduction of the experimental pose and supporting the validity of the docking parameters. In contrast, captopril re-docked into ACE (PDB ID: 1UFZ) yielded an RMSD of 4.87 \AA{}, likely due to the inherent flexibility of the ACE active site and potential limitations of the scoring function for metalloproteins. These results highlight that the docking protocol is generally robust, though predictive performance varies with target--ligand complexity.

\begin{table}[htbp]
\centering
\caption{Docking scores of reference drugs and generated ligands docked into their respective targets, with lower scores indicating stronger predicted binding affinities.}
\label{tab:1}
\begin{tabular}{@{} l r l r @{}}
\toprule
\multicolumn{2}{c}{PPAR$\gamma$} & \multicolumn{2}{c}{ACE} \\
\cmidrule(lr){1-2} \cmidrule(lr){3-4}
Ligand & Docking score & Ligand & Docking score \\
\midrule
Rosiglitazone (redock) & -7.422 & P12821-293 & -9.917 \\
Balsalazide            & -7.399 & P12821-10  & -9.485 \\
P37231-165             & -6.749 & P12821-269 & -9.367 \\
P37231-117             & -6.461 & P12821-225 & -9.29  \\
P37231-39              & -5.464 & Enalapril   & -8.283 \\
P37231-247             & -5.394 & Captopril (redock) & -6.168 \\
\bottomrule
\end{tabular}
\end{table}
% ─── Discussion ─────────────────────────────────────────────────────────────
\section{Discussion}

In this study, we present DrugGen-2, a novel disease-aware generative model that advances disease-conditioned molecular design by generating small molecules tailored to both disease ontologies and target protein sequences. Compared with previously developed models, DrugGen-2 demonstrates improved generation capability, producing novel molecules with higher binding affinities and greater structural similarity to approved drugs. These advances demonstrate DrugGen-2's potential to transform disease-aware drug discovery, bridging the gap between molecular design and translational pharmacology.
DrugGen-2 includes disease context by conditioning on MeSH ontology hierarchies, which guide molecular generation to better align with disease biology. For instance, in DN, where multiple pathways converge on targets like PPAR$\gamma$ or ACE, the model's ontology-based conditioning enables it to generate ligands specified for a target in desired pathways. For instance, while ACE promotes angiotensin-2 formation in the cardiovascular system--leading to hypertension and vascular inflammation-- it simultaneously degrades $\beta$-amyloid peptides in the brain, exerting neuroprotective effects ~\cite{ref22,ref23}. By capturing such context-dependent molecular functions, DrugGen-2 can design compounds that account for disease-specific complexities absent in earlier target-based models. This is evidenced by the superior unique molecule generation rates (up to 444 out of 500 attempts) and binding affinities (median pKd >9.0) across all evaluated targets, outperforming baselines like DrugGPT and DrugGen.
Our results demonstrated DrugGen-2's superior performance in generating unique molecules, reflecting an enhanced exploratory capacity in chemical space, an essential step in early-stage drug discovery, where structural diversity increases the likelihood of finding viable leads ~\cite{ref24}. Across MeSH encodings, its variants consistently produced over 400 unique candidates per target, outperforming DrugGPT (median 219) and DrugGen (median 50). This results from GRPO integration with reward functions prioritizing novelty and diversity, reducing repetitive outputs. Additionally, near-perfect validity rates (99--100\%) confirm DrugGen-2's ability to maintain chemical feasibility and structural integrity during generation. The model's generation of molecules with high similarity to approved drugs (median 0.70 across MeSH variants) and superior predicted binding affinities (medians 9.26--9.97) highlights its focus on clinically relevant molecular scaffolds. Docking simulations further supported these findings, with several ligands demonstrating stronger ACE binding than references like Enalapril, despite minor RMSD variations in flexible binding pockets. Together, the reinforcement learning framework using PLAPT-derived binding rewards guides output toward potent, diverse, and pharmacologically relevant candidates. This gives DrugGen-2 an advantage over DrugGPT and DrugGen in speeding hit-to-lead processes while minimizing risks associated with unexplored chemical entities.
Despite these advances, limitations exist. While PLAPT predictions and GLIDE docking provide in silico insights, experimental validation--such as in vitro binding assays ~\cite{ref25} or in vivo models ~\cite{ref26}--is essential to confirm binding affinities, evaluate functional effects, and assess potential off-target interactions. Moreover, the truncation of long protein sequences to 768 tokens may overlook distal domains critical for allosteric modulation ~\cite{ref27}. Finally, the binary novelty reward might undervalue subtle innovations over radical departures from approved scaffolds. Future studies could incorporate advanced rewards, like absorption, distribution, metabolism, excretion, and toxicity (ADMET) predictions ~\cite{ref28} or synthetic accessibility scores ~\cite{ref29}, to further bridge the gap to clinical candidates.
In conclusion, DrugGen-2 represents a significant step toward AI-assisted pharmacology, demonstrating that disease-conditioned generative models can yield high-quality, context-aware drug candidates. By addressing the interplay between diseases and targets, it holds promise for expediting therapeutic development in an era of precision medicine.

% ─── Methods ────────────────────────────────────────────────────────────────
\section{Materials and Methods}

\subsection{Dataset preparation}

We created a curated dataset linking approved drugs with their corresponding targets and associated diseases. First, approved drug--target pairs were extracted from DrugBank database (version: 5.1.10)~\cite{ref30}. Next, disease-target associations were obtained from DisGeNET database (version: 3.12.1)~\cite{ref19}, which enabled the generation of preliminary disease-target-drug strings. To ensure clinical relevance, we incorporated approved drug-disease relationships (Phase IV) from the ChEMBL database (CHEMBL33)~\cite{ref31}. Preliminary strings were retained only if a direct drug-disease relationship was confirmed; otherwise, they were excluded. Following is the detailed information on dataset preparation.

\subsubsection{Drug-target dataset}

We retrieved drug--target information from DrugBank, yielding 1,710 small molecules with annotated human targets. Of these,117 compounds were labeled as withdrawn. After a thorough assessment, 50 were excluded due to safety concerns or adverse effects. Available SMILES representations for the selected molecules (1,634 of 1,660) were obtained from DrugBank, ChEMBL, and ZINC20~\cite{ref32} database. From a total of 2116 related protein targets, 27 were not present in UniProt~\cite{ref33} which were mapped to equivalent UniProt ID using reviewed UniProt ID, identical names, or the basic local alignment search tool (BLAST)~\cite{ref34}. The UniProt ID ``Q5JXX5'' was omitted from the database and therefore deleted from further analysis. Finally, the sequences for 2093 proteins were retrieved using the UniProt application programming interface (API).

\subsubsection{Target-disease dataset}

We extracted target-disease information from DisGeNET. The disease concept unique identifier (CUI) was updated from version 2019AA to 2023AA, and corresponding MeSH terms were retrieved via the unified medical language system (UMLS) terminology API. In total, 608 MeSH terms were mapped to 2042 protein targets.

\subsubsection{Drug-disease dataset}

We retrieved approved drug indications (disease-drug relationships) from ChEMBL. Following curation, 1,299 small molecules and their associated 643 MeSH terms were retained, forming the drug--disease dataset. To capture hierarchical disease relationships, MeSH terms were mapped to their corresponding DAG structures, resulting in 1,632 DAG representations.

\subsection{Data processing}

We adopted the original DrugGPT tokenizer and extended it with three additional special tokens, ``D'', ``P'', and ``L'', added to the previously defined special tokens, i.e., ``startoftext'', ``endoftext'', and ``PAD'', resulting in a total vocabulary of 53086 tokens. We created the strings of MeSH-sequence-SMILES (1,113,539) and filtered the strings that had the direct MeSH--SMILES relationship which resulted in 13,908 strings of MeSH-sequence-SMILES. Each string was tokenized in the following format: <|startoftext|> + <D> + <MeSH DAG> + <P> + <Sequence> + <L> + <SMILES> + <|endoftext|>. Sequences were padded to a length of 768 tokens, and longer sequences were truncated to this length.

\subsection{Model Development}
\subsubsection{Supervised fine-tuning}
Supervised fine-tuning was performed using the SFT trainer module from the transformer reinforcement learning (TRL) library (version: 0.9.4) ~\cite{ref35} on the pre-trained DrugGPT model. Training was performed for 10 epochs on 13,908 tokenized MeSH--sequence--SMILES strings, with the model at epoch 6 selected for subsequent processing. The SFT configuration was as follows: learning rate = $5 \times 10^{-4}$, batch size = 8, linear warmup with 100 steps, gradient accumulation steps = 1, and the AdamW optimizer with a learning rate of $5 \times 10^{-4}$ and epsilon value of $1 \times 10^{-8}$.
\subsubsection{Group relative policy optimization }
We employed GRPO to further refine the supervised fine-tuned model for de novo drug design. GRPO is a reinforcement learning method that eliminates the need for a separate value function by using group-level statistics to compute advantages and guide policy optimization. This makes GRPO more memory- and computation-efficient while maintaining strong performance.

For each prompt or query $q$, GRPO generates a group of $G$ outputs $\{o_i\}_{i=1}^G$, where each output $o_i$ is sampled from the old policy $\pi_{\theta_{\text{old}}}$ and then optimizes the policy model $\pi_\theta$ by maximizing the objective function $\mathcal{J}_{\mathrm{GRPO}}(\theta)$. Each generated output $o_i$ is scored according to a reward system to assess its binding affinity and validity, novelty compared to approved SMILES, and uniqueness of generation in each batch. The average reward across the group is computed and serves as a baseline to calculate an advantage $\hat{A}_{i,t}$ for the $t$-th token in each output $o_i$. The advantage measures how much that token's contribution is compared to the baseline. The objective function optimizes the policy parameters $\theta$ to favor outputs with positive advantage, using a clipped ratio to ensure stable updates (Eq.~1). Additionally, a KL divergence penalty weighted by hyperparameter $\beta$ is included between the new policy $\pi_\theta$ and a reference policy $\pi_{\text{ref}}$, the fine-tuned supervised model, to keep the updated policy close to known behaviors and prevent pathological shifts and large policy deviations (Eq.~2).

\begin{align}
\mathcal{J}_{\mathrm{GRPO}}(\theta) &= \mathbb{E}_{q\sim P(Q), \{o_i\}_{i=1}^G\sim \pi_{\theta_{\text{old}}}(O|q)} \Bigg[ \frac{1}{G} \sum_{i=1}^G \frac{1}{|o_i|} \sum_{t=1}^{|o_i|} \Bigg\{ \min\!\bigg( \frac{\pi_\theta (o_{i,t} \mid q, o_{i,<t})}{\pi_{\theta_{\text{old}}} (o_{i,t} \mid q, o_{i,<t})} \hat{A}_{i,t}, \nonumber \\
&\qquad \operatorname{clip}\!\Big( \frac{\pi_\theta (o_{i,t} \mid q, o_{i,<t})}{\pi_{\theta_{\text{old}}} (o_{i,t} \mid q, o_{i,<t})}, 1-\epsilon, 1+\epsilon \Big) \hat{A}_{i,t} \bigg) - \beta \, D_{\mathrm{KL}}[\pi_\theta \| \pi_{\text{ref}}] \Bigg\} \Bigg] \label{eq:grpo}
\end{align}
where $\epsilon$ and $\beta$ are hyperparameters controlling the clipping range and KL penalty weight, respectively.

The KL divergence term is defined as:
\begin{equation}
D_{\mathrm{KL}}[\pi_\theta \| \pi_{\text{ref}}] = \frac{\pi_{\text{ref}} (o_{i,t} \mid q, o_{i,<t})}{\pi_\theta (o_{i,t} \mid q, o_{i,<t})} - \log\!\left( \frac{\pi_{\text{ref}} (o_{i,t} \mid q, o_{i,<t})}{\pi_\theta (o_{i,t} \mid q, o_{i,<t})} \right) - 1 \label{eq:kl}
\end{equation}

We implemented GRPO using a custom \texttt{GRPOTrainer} class from the Hugging Face Transformers TRL library (version 0.23.1). Training configurations were defined through a \texttt{GRPOConfig} object specifying both optimization and generation parameters.

The optimization was carried out with the AdamW optimizer, using hyperparameters $\beta_1 = 0.9$, $\beta_2 = 0.999$, $\epsilon = 1 \times 10^{-8}$, and a weight decay of $0.1$. The learning rate was initialized at $5 \times 10^{-6}$ and adjusted using a linear learning rate scheduler, with the first 100 steps designated for warm-up. To stabilize updates, we applied gradient accumulation with an effective batch factor of 4, enabled mixed-precision training (fp16), and enforced gradient clipping at a maximum norm of $1.0$. The total training was performed for 10 epochs. Logging and checkpointing were executed at fixed step intervals. For reproducibility, both the global seed and data seed were set to 42.

SMILES generation during training was controlled by a defined set of hyperparameters. The maximum input length was fixed at 768 tokens, while the GRPO group size (number of completions per prompt) was set to 40. The model operated with a context window of 1,024 tokens, and the maximum generation length was limited to 256 tokens (i.e., $1024 - 768$). Sampling was performed with default values of $1.0$ for top $p$ and \texttt{None} for top $k$. The temperature was set to $1.0$ to preserve stochasticity in outputs.

\subsubsection{Validity-checker pipeline}
During GRPO training, we implemented a validity-checker pipeline using the RDKit cheminformatics toolkit (version 2023.9.5)~\cite{ref36} to filter out syntactically or chemically invalid molecular representations. The pipeline applies a series of rule-based checks designed to detect structural inconsistencies. First, each SMILES string is read by RDKit, and strings that cannot be interpreted or do not produce a valid molecular graph are flagged as invalid. Molecules containing fewer than two atoms are also rejected, as such structures are considered chemically trivial. Valence state consistency is then examined, which enforces chemically permitted bonding rules; molecules that contain atoms with impossible valence states are classified as invalid. The results of these validation steps are integrated into a unified system, whereas molecules passing all checks are classified as chemically valid.
\subsubsection{Reward functions}
During training, three distinct reward functions were employed to guide optimization toward generating molecules with desirable properties: maximizing predicted binding affinity while penalizing invalid structures, ensuring novelty relative to approved drugs, and encouraging diversity by preventing repetition within each batch.

\begin{itemize}
    \item Binding affinity: To measure the binding affinity, we used PLAPT, which is a deep learning model developed to measure the binding affinity between protein and SMILES. This reward function computes reward values for batches of generated molecular sequences by evaluating candidate SMILES strings against corresponding target proteins. For each input--completion pair, the function first checks the validity of the generated SMILES using the validity-checker pipeline. Invalid SMILES are assigned a reward of zero. If the SMILES is valid, the function extracts the protein sequence embedded within the input prompt, specifically the segment located between the \texttt{<P>} and \texttt{<L>} tags. This extracted sequence, together with the candidate SMILES, is then passed to the PLAPT model, which returns a predicted binding affinity score defined as the negative logarithm (base 10) of the dissociation constant ($K_d$). The function ultimately returns a list of rewards corresponding to each input pair of prompt and completion.

    \item Similarity to approved molecules: To promote novelty, each generated molecule was compared against our dataset of approved drugs. A binary reward was assigned such that molecules already present in the dataset received a reward of zero, whereas novel molecules absent from the dataset received a reward of one.

    \item Avoidance of repeated molecules within a batch: To discourage reward hacking through duplication, we implemented a batch-level reward. Each generated SMILES was assigned a binary reward of one if it was unique within the batch, or zero if it appeared multiple times in the same batch. This ensured diversity among the generated candidates during training.
\end{itemize}

\subsection{Assessment}
To evaluate the capabilities of DrugGen-2 in generating high-quality small molecules, we conducted a series of experiments based on selecting diabetic nephropathy as a representative example. Initially, we identified DN-related targets from DisGeNET and filtered five highly druggable target proteins using the DrugTar algorithm: ACE, PPAR$\gamma$, NOS3, PAI-1, and TGF-$\beta$1. For model input preparation, DrugGen-2 incorporates disease information through MeSH DAG structures paired with target proteins. In the case of DN, the MeSH descriptor ``Diabetic Nephropathies'' was mapped to the following hierarchical identifiers:
\begin{itemize}
    \item C12.050.351.968.419.192: Urogenital Diseases, Female Urogenital Diseases and Pregnancy Complications, Female Urogenital Diseases, Urologic Diseases, Kidney Diseases, Diabetic Nephropathies
    \item C12.200.777.419.192: Urogenital Diseases, Male Urogenital Diseases, Urologic Diseases, Kidney Diseases, Diabetic Nephropathies
    \item C12.950.419.192: Urogenital Diseases, Urologic Diseases, Kidney Diseases, Diabetic Nephropathies
    \item C19.246.099.875: Endocrine System Diseases, Diabetes Mellitus, Diabetes Complications, Diabetic Nephropathies
\end{itemize}
We then compared the performance of DrugGen-2 against DrugGen and DrugGPT models with their default parameters across five key evaluation metrics: the capability of generating unique small molecules, the validity of the generated molecules, the similarity of these molecules to approved drugs, and their binding affinity to the selected targets. Finally, to provide deeper insights into DrugGen-2's outputs, we conducted docking simulations for further analysis.
\subsubsection{Generation power}
To assess the models' capability to generate unique small molecules, we directed each model to produce 500 unique small molecules per input. For DrugGPT and DrugGen, inputs consisted of the five selected DN-associated target proteins, whereas DrugGen-2 was queried with MeSH DAG--target protein pairs. If a model failed to generate novel molecules within 30 sampling cycles, the generation process for that input was terminated. The number of successfully generated unique molecules was taken as a measure of generative capability. Statistical analysis was conducted using the $\chi^2$ test, and correction for multiple comparisons was performed using the Benjamini--Hochberg method.
\subsubsection{Validity assessment }
For each DrugGen-2 input pair (MeSH DAG--target protein) and for each target protein input of DrugGen and DrugGPT, the models were instructed to generate 100 unique molecules. The outputs were filtered using the validity-checker pipeline described earlier, and the percentage of valid structures was reported as a measure of each model's ability to generate chemically valid molecules. Statistical analysis was conducted using the $\chi^2$ test.
\subsubsection{Similarity to the approved small molecules}
To evaluate the similarity of generated molecules to approved drugs, we used the same set of outputs from the validity assessment. Each valid molecule, as determined by the RDKit pipeline, was converted into a Morgan fingerprint (radius = 2, size = 2048 bits). Pairwise Tanimoto similarity scores were computed between the fingerprints of generated and approved SMILES, and the maximum similarity value was reported as the final similarity score. Statistical analysis was conducted using the Kruskal--Wallis test, and corrections for multiple comparisons were applied using the Benjamini--Hochberg method.
\subsubsection{Binding affinity using PLAPT}
The same set of molecules generated during the validity assessment was used to evaluate the binding affinities of the compounds produced by DrugGen-2, DrugGen, and DrugGPT. The invalid structures were considered in the binding affinity assessments. Statistical analysis was conducted using the Kruskal-Wallis test, and corrections for multiple comparisons were applied using the Benjamini-Hochberg method.
\subsubsection{Docking simulation}
To investigate binding interactions, molecular docking was performed against two proteins with available crystal structures from the Protein Data Bank (PDB): ACE (UniProt ID: P12821 and PDB ID: 1UFZ) and PPAR$\gamma$ (UniProt ID: P37231 and PDB ID: 4EMA). These targets were chosen for their therapeutic relevance, as well as the availability of co-crystallized ligands, which allowed validation of the docking protocol. After removing duplicate entries, a curated library of newly designed ligands was docked using a blind docking approach. For benchmarking, at least one approved small-molecule drug per target was included, retrieved from DrugBank. Docking grids were uniformly set to $40 \times 40 \times 40$~\AA$^3$, the maximum permissible by the software, to ensure comprehensive exploration of potential binding regions~\cite{ref37}.

Protein structures were prepared using Schr\"{o}dinger's Protein Preparation Wizard~\cite{ref38}, including hydrogen addition, optimization of hydrogen-bonding networks, and assignment of protonation states at physiological pH (7.4 $\pm$ 0.5). Ligands were processed with LigPrep~\cite{ref39} under the OPLS4 force field~\cite{ref40}, generating all relevant ionization and tautomeric states.

Docking simulations were carried out in Maestro (v2021-2) using the GLIDE XP algorithm~\cite{ref41}. Ligands were treated as flexible, while proteins were kept rigid. Reliability of the docking workflow was assessed by re-docking the original co-crystallized ligands into their respective protein structures. Docking outcomes were ranked by GLIDE XP scores, where more negative values correspond to stronger predicted binding affinities. Complete docking data and SMILES representations of the ligands are available in Supporting Information 4.
% ─── Data & Code Availability ───────────────────────────────────────────────
\newpage
\section*{Data availability}

All data generated or analyzed during this study are included in the manuscript and supporting information files. The MeSH-sequence-SMILES dataset of approved disease-target-drug pairs used in this study is publicly available at \url{https://huggingface.co/datasets/alimotahharynia/approved_disease_target_drug}.

\section*{Code availability}

The checkpoints and code for generating small molecules are publicly available at \url{https://huggingface.co/alimotahharynia/DrugGen-2} and \url{https://github.com/alimotahharynia/DrugGen-2}, resoectively. Explore the interactive user interface of DrugGen-2 at \url{https://huggingface.co/spaces/alimotahharynia/DrugGen-2}.

% ─── Acknowledgments ────────────────────────────────────────────────────────
\section*{Acknowledgments}

We sincerely thank Dr. Farnoush Kiyanpour for her valuable comments on the first draft of the manuscript.

% ─── Funding ────────────────────────────────────────────────────────────────
\section*{Funding}

No funding was received for this study or its publication.

% ─── Competing interests ────────────────────────────────────────────────────
\section*{Competing interests}

The authors declare no competing interests.

% ─── Author contributions ───────────────────────────────────────────────────
\section*{Author contributions}

Conceptualization: A.M, Y.G. Dataset preparation: A.M, M.S, N.M. Model development: A.M, M.G, N.M, M.I. Model assessment: A.M, M.G. Statistical analysis: A.M, M.G. Molecular docking: M.S, H.S. Data interpretation: All authors. Drafting original manuscript: A.M, M.G, M.S, N.M. Revising the manuscript: A.M, M.I, H.S, Y.G. All the authors have read and approved the final version for publication and agreed to be responsible for the integrity of the study.

% ─── References ─────────────────────────────────────────────────────────────

\end{document}